\documentclass[aps,prl,reprint,showpacs,floatfix,groupedaddress]{revtex4-1}
\usepackage{multirow}
\usepackage[T1]{fontenc}
\usepackage[cp1250]{inputenc}
\usepackage{psfrag}
\usepackage{graphicx}
\usepackage[english]{babel}
\newcommand{\bq}{\begin{equation}}
\newcommand{\eq}{\end{equation}}
\newcommand{\ba}{\begin{eqnarray}}
\newcommand{\ea}{\end{eqnarray}}
\begin{document}
\title{Emerging long orbits and self-similar temporal sequences in classical oscillators}
\author{Darka Labavi\'c}
\author{Hildegard Meyer-Ortmanns}
\email{h.ortmanns@jacobs-university.de}
\affiliation{Physics and Earth Sciences, Jacobs University, P.O. Box 750561, 28725 Bremen, Germany}

\noindent
\begin{abstract}
We analyze repulsively coupled Kuramoto oscillators, which are exposed to a distribution of natural frequencies. This source of disorder leads to closed orbits with a variety of different periods, which can be  orders of magnitude longer than periods of individual oscillators. By construction the attractor space is quite rich. This may cause long transients until the deterministic trajectories find their stationary orbits. The smaller the width of the distribution about the common natural frequency is, the longer are the emerging time scales on average. Among the long-period orbits we find  self-similar sequences of temporary phase-locked motion on different time scales. The ratio of time scales is determined by the ratio of widths of the distributions.
\end{abstract}


\maketitle
As model for classical oscillators we consider the Kuramoto model, a paradigm for synchronization phenomena with applications to natural and artificial systems \cite{acebron}. While in earlier times modifications of the Kuramoto model were mostly considered for attractive (positive) couplings between the oscillators, nowadays negative couplings are also discussed in the context of neural networks, representing inhibition \cite{ledoux}, or social science, representing contrarians \cite{hong}. When negative couplings come into the game, a notion of frustrated bonds can be defined, going back to \cite{daido} and later considered in \cite{pablo}. As it turns out, frustration in oscillatory and excitable systems can lead to analogous effects as in spin glasses. The first one to mention is multistable behavior, where the attractors are much more versatile in oscillatory systems than in spin glasses. Also order-by-disorder phenomena can be identified \cite{misha}, where in replacement of thermal fluctuations in spin glasses, disorder in oscillatory systems is realized as additive or multiplicative noise.

While first hints to a particularly rich attractor space for certain grid sizes of repulsively coupled oscillators were identified in \cite{misha}, a closer exploration of the attractor spaces in \cite{tomov,zaks} showed what we would call a ``dynamically generated Watanabe-Strogatz phenomenon". Starting from a considerable fraction of initial conditions and uniform natural frequencies, the oscillator phases arrange themselves into clusters of almost coalescing phases, where the clusters, considered as collective variables,  satisfy the criteria for  dimensional reduction to happen as predicted by Watanabe and Strogatz \cite{watanabe}. This phenomenon goes along with N-3 conserved quantities and continuous sets of solutions.
This way the mechanism realizes a case of so-called extreme multistability \cite{feudel} with a continuum of attractors. Conditions, under which more generally this interesting phenomenon can be observed, were explored by M. Zaks \cite{zaks}. Effects based on multistability should be particularly pronounced, when such a  continuum of attractors is available. Therefore we shall choose our system of oscillators with respect to size, geometry and coupling accordingly \cite{florin} and slightly perturb about the uniform natural frequency distribution. Later we compare the effects  with cases of moderate multistability. As we shall see, the time evolution of oscillator phases explores these attractor spaces quite differently from a migration under the action of noise \cite{florin}. This paper will merely focus on a phenomenological description of the time evolution of trajectories with amazing features of long periods and self-similar temporal patterns.

\begin{figure}
	\begin{center}
		\includegraphics[width=8cm]{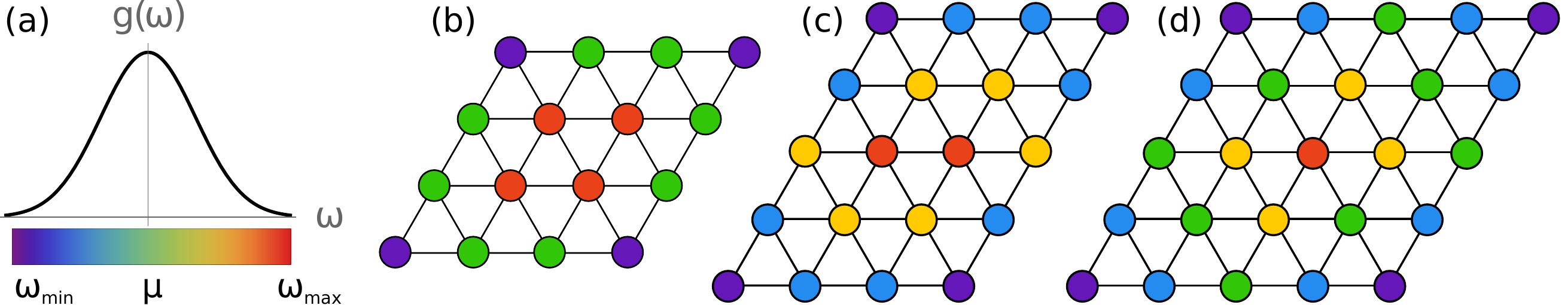}
		\end{center}
	\caption{Regular assignment of natural frequencies. Three different sizes $l_x\times l_y = 4\times4, 5\times4, 5\times5$ of a hexagonal lattice. Colors represent a value of $\omega$, orange being the largest, and purple the smallest.
	}
\label{fig1}
\end{figure}

We consider systems of $N$ Kuramoto oscillators~\cite{kuramoto},
whose phases $\phi_i$ are governed by the equations:
\begin{equation}
\label{eq1}
\frac{d\phi_i}{dt} =
\omega_i +  \frac{\kappa}{\mathcal{N}_i}\sum_j A_{ij}\sin(\phi_j-\phi_i).
\end{equation}
Here $\omega_i$ denote the natural frequencies,
$\kappa$ parameterizes the coupling strength, chosen to be $\kappa=-2$ throughout the paper, that is deeply in the regime of multistability. Further, $\mathcal{N}_i$ denotes the number of neighbors to which the $i$-th unit is connected, and $A_{ij}$
is the adjacency matrix with $A_{ii}=0$, $A_{ij}=1$ if
$i\neq j$ and units $i$ and $j$ are connected, otherwise $A_{ij}=0$.
We choose $A_{ij}$ to represent a hexagonal lattice of the size
$L\times L$ with periodic
boundary conditions, so that $\mathcal{N}_i=6$ for all sites i. For the frequency distributions we choose a regular and a random implementation. For the random version we assume that the number density of natural frequencies follows, on average, the Gaussian distribution $g(\omega)=\frac{1}{\sqrt{2\pi\sigma^2}}e^{-(\omega-\mu)^2/(2\sigma^2)}$, describing frequency fluctuations about $\mu=1$ with typically $\sigma=0.01$.
For the regular implementation we generate frequencies according to a deterministic procedure. The analytic expressions can be found in \cite{suppl}. The choice is most easily illustrated in Fig.~\ref{fig1}.
The frequencies are uniformly spaced with respect to $g(\omega)$ and generate ``spatial" Gaussians in natural frequencies with the largest frequencies in the center of the grid (e.g., four sites in the middle of the $4\times4$ grid of Fig.~\ref{fig1}). This way the deterministic system retains some symmetry in the frequency distribution over the grid and can be smoothly tuned towards the uniform frequency. We shall analyze the dependence on the initial conditions for $\phi_i$, keeping the distribution of $\omega_j$ fixed.

\begin{figure}
	\begin{center}
		\includegraphics[width=8cm]{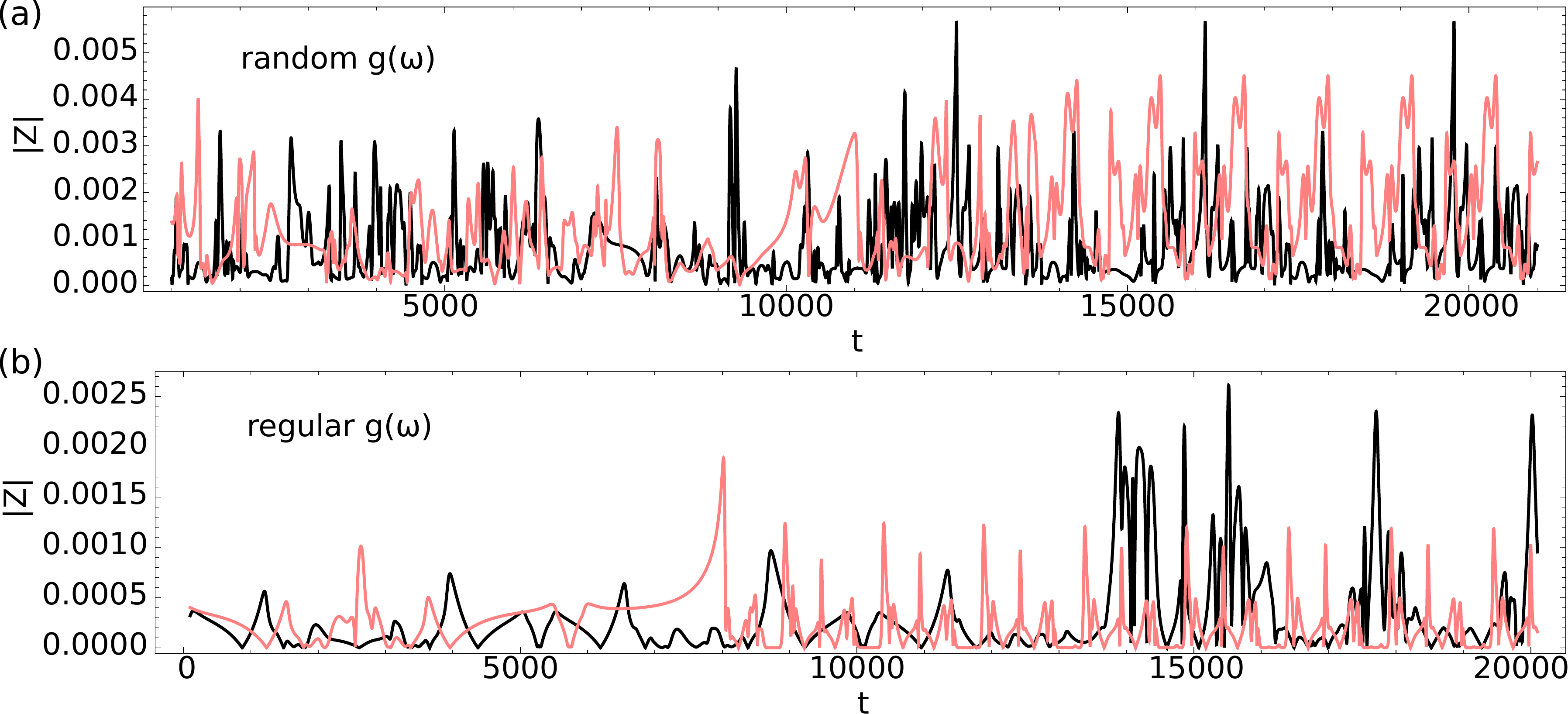}
		\end{center}
	\caption{Transients of the order parameter for initial conditions around the fixed point. Two trajectories in black and red are shown for a $4\times4$ hexagonal lattice for random $\omega$'s, corresponding to two different realizations of $g(\omega)$ with the same mean $\mu$ and standard deviation $\sigma$, and regular $\omega$'s, corresponding to two different initial conditions, respectively. All initial conditions are taken from a small radius around $\varphi_i=0.5\;\forall i$. The parameters are $\mu=1$, $\sigma=0.01$, and $\kappa=-2$.  }
\label{fig3}
\end{figure}

To project on the interesting features it turns out to be quite convenient to study first the time evolution of the Kuramoto order parameter, rather than the individual trajectories of phase-correlated motion. It is defined as $Z(t):= \vert Z\vert e^{i\theta}\;=\;\frac{1}{N}\sum_{j=1}^N e^{i\phi_j}$,
where a positive value of $\vert Z\vert$ in the limit of $N\rightarrow\infty$ implies the emergence of phase synchronization and $\theta$ is the phase of the global order parameter \cite{kuramoto}.
It is first $\vert Z\vert$, for which we observe striking long periods, by orders of magnitude longer than individual periods of oscillators. Although identical values of $\vert Z\vert$ are compatible with, but not conclusive for the same patterns of phase-locked motion,  it is a useful first tool to record and register these long periods. Long periods of the trajectories may be easily overlooked when following the individual time evolution of larger sets of oscillators. As a next step we then focus on selected points of the order parameter and analyze the trajectories of individual oscillators to identify their period and synchronization pattern. As we shall see, apart from constant values for $\vert Z\vert$, corresponding to stable limit cycles with constant frequencies and periods of the order of individual cycles, we observe long transients to reach a stable orbit and long periods of these orbits, for both types of frequency distributions and both kind of initial conditions. As initial conditions we either start from values close to the fixed point (in phase differences, which is a unique solution for positive $\kappa$ and far off the desired phase pattern for negative $\kappa$) (as we used for Fig.~\ref{fig3}), or instead, from a uniform distribution in the interval $(0, 2\pi)$, displayed in \cite{suppl}. All presented order parameters are periodic orbits with different maxima and different periods. Transients are of the same order of magnitude, ranging from $\sim10000$ (bottom gray (online red) trajectory in Fig.~\ref{fig3}, random assignment, to $\sim 17000$ t.u. (bottom black curve, Fig.~\ref{fig3}, regular assignment)  (and similarly from $\sim10000$ to $\sim 16000$ for the second type of initial conditions, shown in \cite{suppl}. For Fig.~\ref{fig3} we pursued the periodic behavior until 500000 time units (t.u.).

Unless the choice of initial conditions would accidentally hit the vicinity of an attractor for $\kappa<0$, transients can take rather long, independently of whether starting from the neighborhood of a distinguished value in the monostable regime ($\kappa>0$, which is certainly far off from the finally approached phase pattern), or from a uniform distribution between 0 and $2\pi$, so that the long duration depends on the attractor space itself, as we shall later confirm by a comparison to results for the $5\times5$-lattice. The following  Fig.~\ref{fig5a}  displays the periodic time evolution of $\vert Z\vert$ once it has stabilized after the transients. Depending on the realization of $\omega_i$'s and/or initial conditions, the system evolves to a different attractor. Attractors here are characterized by means of the time evolution of $\vert Z\vert$ that indicates different temporary synchronization patterns during the orbit, before we zoom into individual trajectories. All four order parameters in Fig.~\ref{fig5a} have different periodic orbits.

\begin{figure}
	\begin{center}
		\includegraphics[width=8cm]{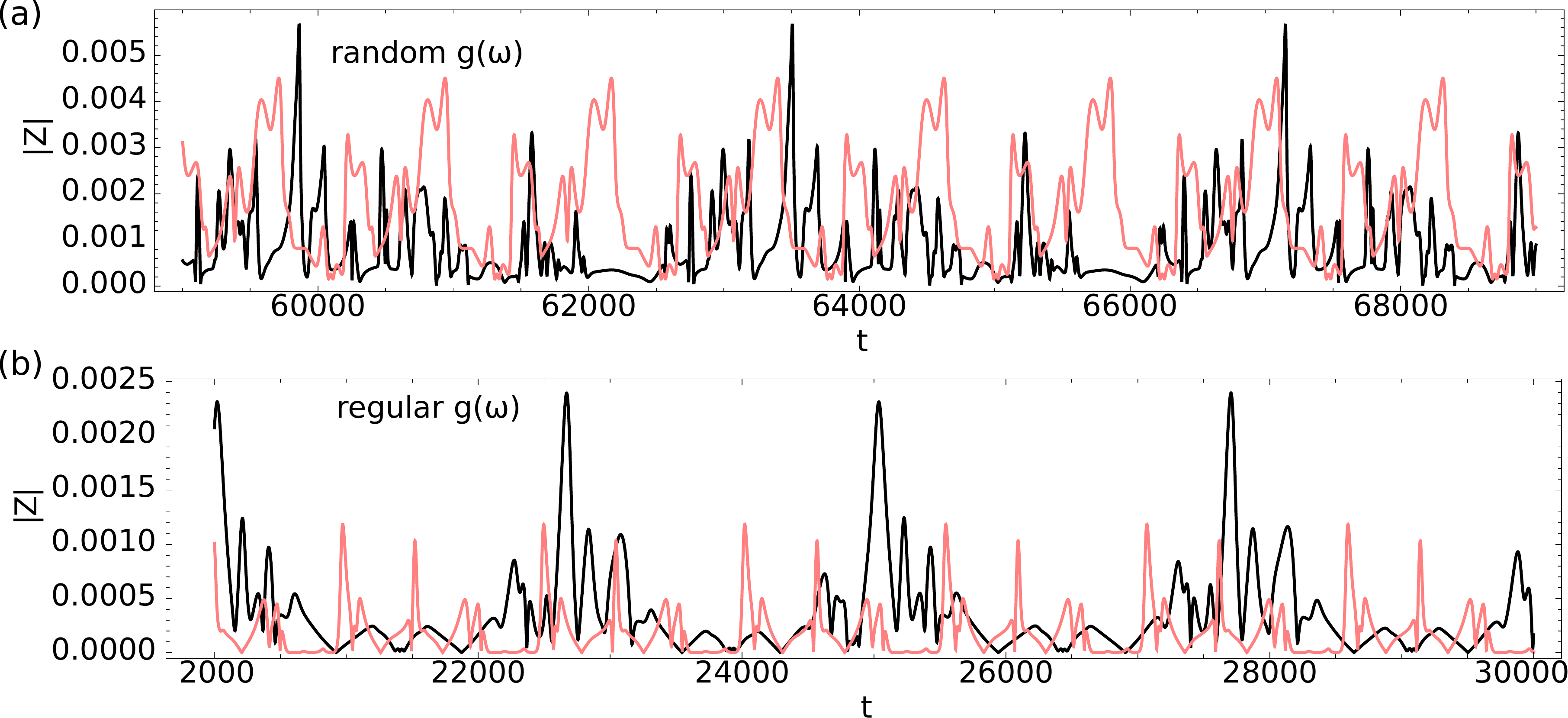}
		\end{center}
	\caption{Periodic time evolution of the order parameter with initial conditions around the fixed point, for the same trajectories of Fig~\ref{fig3}.}
\label{fig5a}
\end{figure}

The time evolution of the $\vert Z\vert$ with initial conditions from $\varphi_i\in(0,2\pi)$ show periods ranging from 1000 to 10000 for both random and regular $g(\omega)$.

\begin{figure}
	\begin{center}
		\includegraphics[width=7cm]{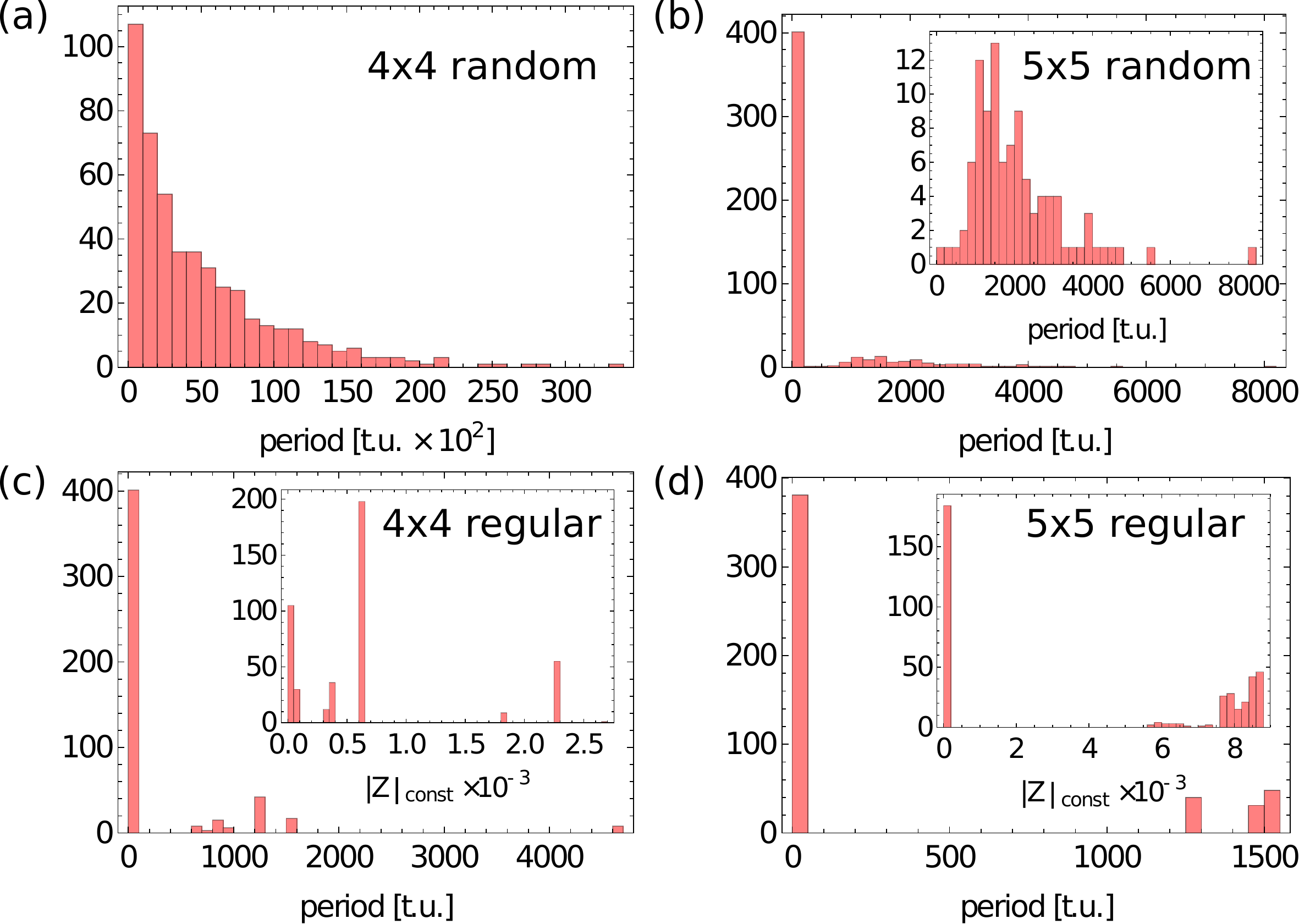}
		\end{center}
	\caption{
	Histograms of order parameter periods as a function of the period length for 500 realizations of $g(\omega)$ for random $\omega$'s, and 500 different initial conditions for regular $\omega$'s for two different lattice sizes. Parameters are $\mu=1$, $\sigma=0.01$, $\kappa=-2$, initial conditions are taken from a vicinity of the fixed point at $\varphi_i = 0.5$.}
\label{fig6}
\end{figure}
In order to get hints on how representative the long-period orbits and the long transients are for the systems we consider, we display in Fig.~\ref{fig6} histograms of the order parameter
periods for both lattice sizes and frequency implementations. An order parameter with zero period corresponds to a constant order parameter.  We see a clear difference between the two lattice sizes for random distribution of $\omega$'s. In both cases there is a clear peak at zero, for  the $4\times4$ lattice 17\% of the realizations (for a fixed initial condition) evolves to an attractor with a constant order parameter, compared to 80\% in the case for the  $5\times5$ lattice. Beside this larger number of zero-period order parameter orbits (that is, limit cycles), the periods are also longer for $4\times 4$ by an order of magnitude. Similarly to the $5\times5$-grid, also the $4\times5$-grid (not displayed) has a large number of zero-period orbits, but a few long-period orbits as well, whose periods are in general shorter than in the $4\times4$ case, but still orders of magnitude longer than the individual oscillation period.

In the case of regular $\omega$'s, the difference is less pronounced for $4\times4$ and $5\times5$ lattices: Most of the initial conditions lead to a constant order parameter, and we find only a few different periodic orbits. For a zero-period order parameter, i.e. a constant, for the regular assignment of $\omega$'s, we show in the insets of (c) and (d) how many different constant order parameters we find, corresponding to different limit cycles for different initial conditions, and indicating multistability. For random assignments, different order parameters, recorded for different realizations, are not conclusive in view of multistability, as the systems differ by their natural frequencies. So far we changed initial conditions only for one random realization and found bistable behavior: two different order parameters, one constant, and one periodic.

In this kind of systems a complete overview of the possible states is beyond the scope of this paper.	 More insight into the versatility of multistability was obtained for uniform frequency distributions in \cite{tomov,zaks}, and our frequency distributions are perturbations about the uniform cases considered there. From the results of  \cite{tomov,zaks} a clear difference between the attractor spaces for these two lattice sizes can be traced back to a larger number of possible 4-cluster partitions  for the $4\times 4$ case, which can be assigned to the grid as patterns of phase-locked motion. The lattice allows the phases to organize themselves into four actually globally coupled clusters, each of four coalescing phases. It is this system of emerging four globally coupled clusters that satisfies the conditions for dimensional reduction \`a la Watanabe and Strogatz \cite{watanabe}, therefore leading to a continuum of attractors, with one conserved quantity (see below), and a continuum of solutions differing by frequencies. Depending on the width of our frequency distributions about $\mu$,  we violate the conditions for the occurrence of the ``Watanabe-Strogatz phenomenon" more or less slightly, so we expect to see ``remnants" of the different attractor spaces at uniform frequencies for different widths, when we turn on deviations from the uniform natural frequency. Before we discuss the dependence of the transient times and periods on the strength of the deviations, which reveal some remnants, let us take a closer look at the long-period orbits.

When assigning phases to the unit cycle, each lattice site is characterized by a gray scale (color online) and a size of a disk, so that in principle we can disentangle sixteen even coalescing phases on the unit cycle as a sphere with 16 non-overlapping internal rings, see \cite{suppl}.

\begin{figure}
	\begin{center}
		\includegraphics[width=8cm]{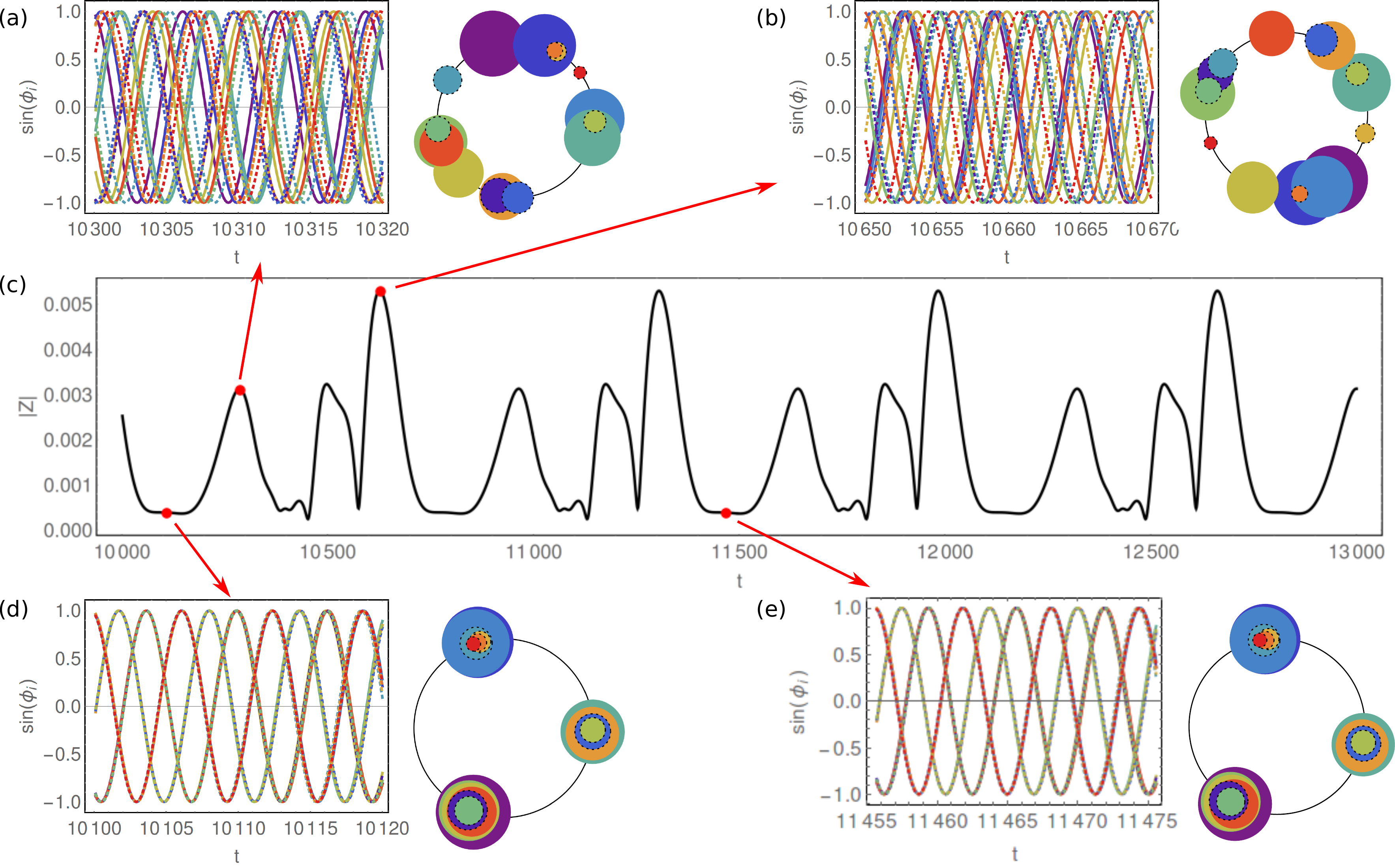}
		\end{center}
	\caption{Time evolution of the phases and order parameter, phase snapshots on the unit cycle for one choice of initial conditions for regular $\omega$'s. The parameters are $\mu=1$, $\sigma=0.01$, $\kappa=-2$ as before with initial conditions taken from a vicinity of the fixed point at $\varphi_i = 0.5$.}
\label{fig8}
\end{figure}

In order to understand what the long-period orbits of the Kuramoto order parameter mean for the trajectories, we show the time evolution of $\vert Z\vert$ for one choice of initial conditions for regular $\omega$'s. For selected instants of time, indicated by the gray (online red) points in Fig.~\ref{fig8}, we show a snapshot of the sixteen phases together with their representation on the unit cycle. While the time evolution of the phases indicate a phase period of about 6 t.u. $\approx 2\pi/\omega_i$, the total period of the periodic orbit of $\approx 1500$ t.u. is  about 100 times longer in this case. Moreover, in the concrete case of Fig.~\ref{fig8}, only every second absolute maximum corresponds to the same state, because two neighboring absolute maxima correspond to different phase patterns, which add up to the same value of $\vert Z\vert$. For all long-period orbits in our figures we have checked when the return of the value of $\vert Z\vert$ meant the same pattern of phase-locked motion, so that the long periods of $\vert Z\vert$ are actually representative for long periods of the system. As such an example the bottom row of Fig.~\ref{fig8} shows two (almost) the same states of the system, a three cluster state (with three bubbles along the unit cycle). It is numerically hard to exactly hit the instant of time, when states like the three-cluster state are exactly the same, as they are temporary patterns. In the movie (see \cite{suppl}) we show the time evolution over one whole period, that is 272 seconds, corresponding to 1355.8 t.u. or 13558 frames in the movie, so that the periodicity in configurations of phases becomes visible if one is patient to wait until the end. 

A closer look at the evolution of phase patterns over a long time such as a whole period reveals approximate temporary n-cluster states, where n varies in time from 2 to 16. From our former analysis \cite{florin} for uniform frequencies we know that a 3-cluster state was unstable, while 4-or 6-clusters states were stable as long as noise was absent to make these states metastable and kick them out of their basin of attraction. The dwell times close to the approximate n-cluster states in the present simulations are not much longer than time intervals, over which phase patterns cannot be uniquely classified in terms of n-cluster states due to the spreading between the phases. So far our observations are compatible with the interpretation that the long-period orbits are heteroclinic cycles, connecting unstable limit cycles. We do not observe longer and longer dwell times when the trajectory comes close to a certain temporary pattern when repeating the cycles; longer dwell times were expected, if these patterns would correspond to saddle points connected by a heteroclinic cycle. As the phase space is sixteen-dimensional, an analytical proof of this interpretation is out of the scope of our phenomenological description. However, in forthcoming work we will analyze the stability of the long-period orbits under variations in the initial conditions and under the action of noise.

\begin{figure}
	\begin{center}
		\includegraphics[width=6cm]{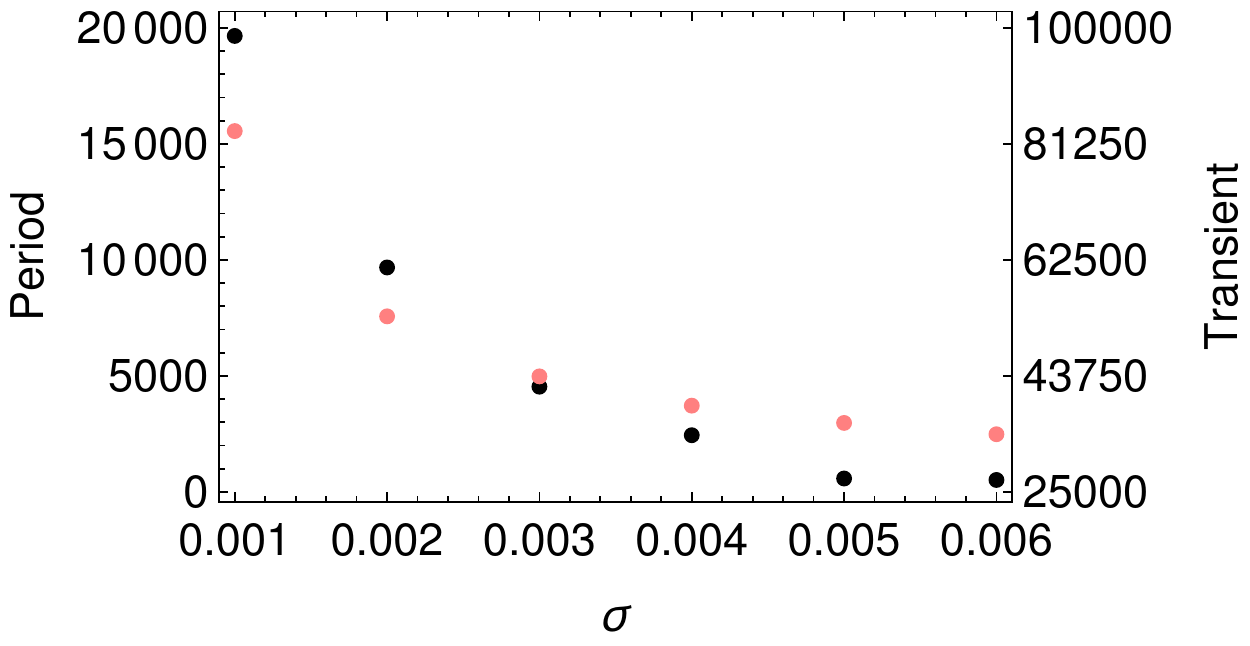}
		\end{center}
	\caption{Transient time and period as a function of the width $\sigma$ of the Gaussian distribution $g(\omega)$ for regular assignment of natural frequencies, fixed $\mu=1$, $\kappa=-2$ and identical initial conditions. Periods (red dots) and transient times (black dots) are decreasing with an increase of the Gaussian width.
	}
\label{fig9}
\end{figure}

\begin{figure}
	\begin{center}
		\includegraphics[width=8cm]{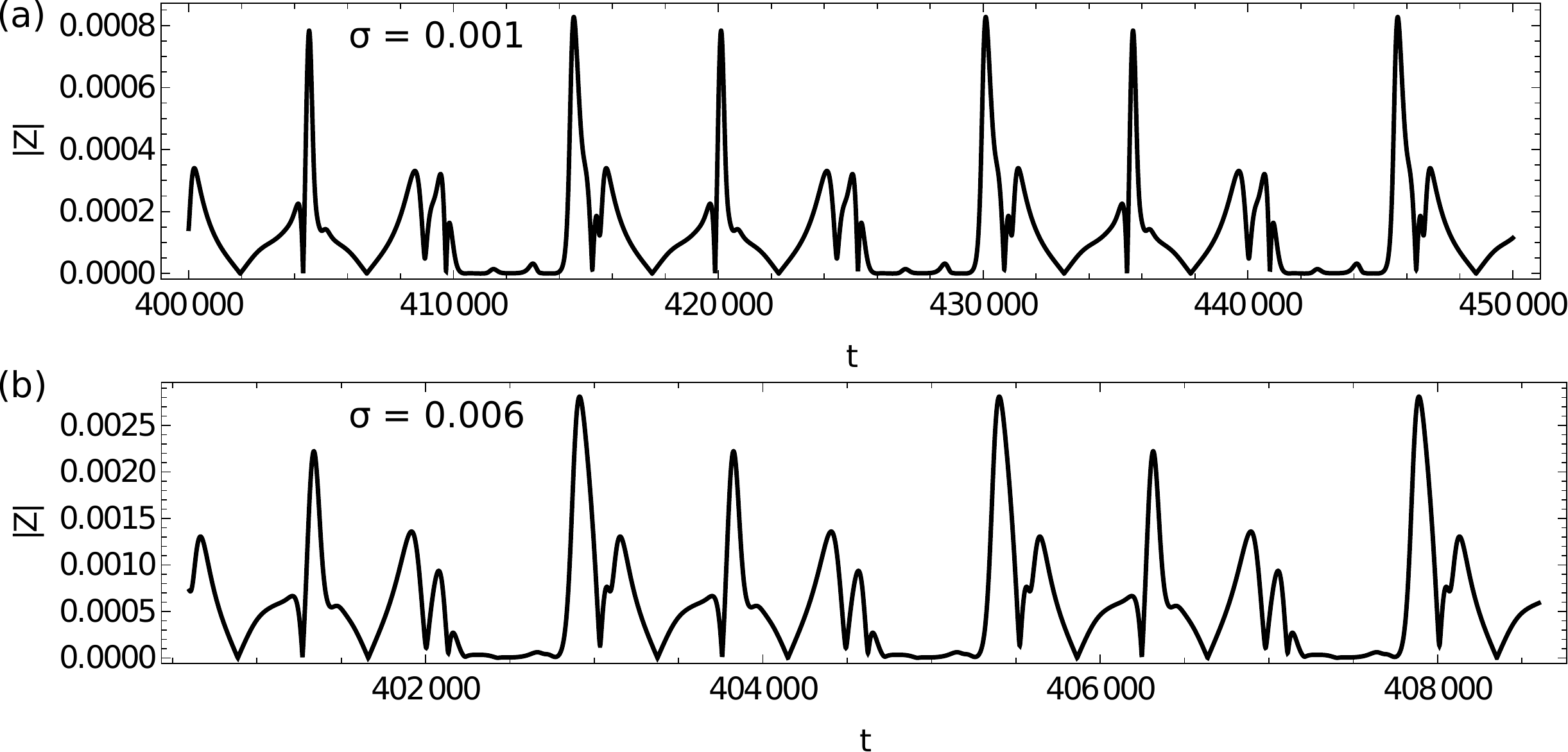}	
		\end{center}
	\caption{Self-similar trajectories of order parameters in time for two different widths of $g(\omega)$. Notice the different time scales. The overall time interval is some 40000 t.u. shorter in the bottom figure.
	}
\label{fig10}
\end{figure}

To support our hypothesis that it is the rich attractor space, which causes long transients towards finding a stable trajectory and a long time to close an orbit (possibly heteroclinic) between unstable limit cycles, we analyze in Fig.~\ref{fig9} the dependence of transient time and period on the width of the Gaussian distribution for a $4\times4$-grid: a tiny width means an almost uniform frequency distribution with a known rather rich attractor space, which should get reduced for larger deviations. To have comparable starting configurations, we choose the same initial conditions for regular frequency assignment and change only the standard deviation $\sigma$ from 0.001 to 0.006, while the other parameters are kept fixed. Even starting with the same initial condition does not guarantee that the system will evolve to the same type of attractor for  different $\sigma$, just in a different transient time. To consider trajectories, which are comparable in both the transient time and the ``attractor type" itself, we make sure that the system actually evolves to the same type in the sense that the periodic orbit has the same shape as illustrated in Fig.~\ref{fig10}, and compare time scales only for such trajectories.
In conclusion to Fig.~\ref{fig9}, it confirms our conjecture that the smaller the width, so the richer the space of attractors for the $4\times 4$ grid is, the longer are the transient times and periods.

In Fig.~\ref{fig10} we show order parameter trajectories for the largest ($\sigma=0.006$) and  smallest ($\sigma=0.001$) Gaussian widths, considered before in Fig.~\ref{fig9}, to illustrate what we mean by the same type of attractor. It should be noticed that the very fact is quite striking to observe this self-similarity of trajectories over time scales, which roughly differ by a factor of 6 (the same factor as given by the ratio of widths). While the period is about 12500 t.u. for $\sigma=0.001$, it is about 2000 t.u. for $\sigma=0.006$. The same scaling of ratios of periods with ratios of widths is observed for other widths as well, see \cite{suppl}. A similar shape of the trajectories means a similar sequence of synchronization patterns. Yet, the absolute values of $\vert Z\vert$ differ at comparable peaks. A zoom into the actual synchronization pattern (not displayed) at time instants of corresponding maxima showed that the temporary states could indeed be classified by n-cluster states, for example three clusters, but the tolerance interval, within which the phases coalesce within a cluster, varies  between both, the widths, and the individual phases. So the self-similarity refers to temporal sequences of synchronization patterns within adapted tolerance intervals of what makes up a cluster.

\begin{figure}
	\begin{center}
		\includegraphics[width=8cm]{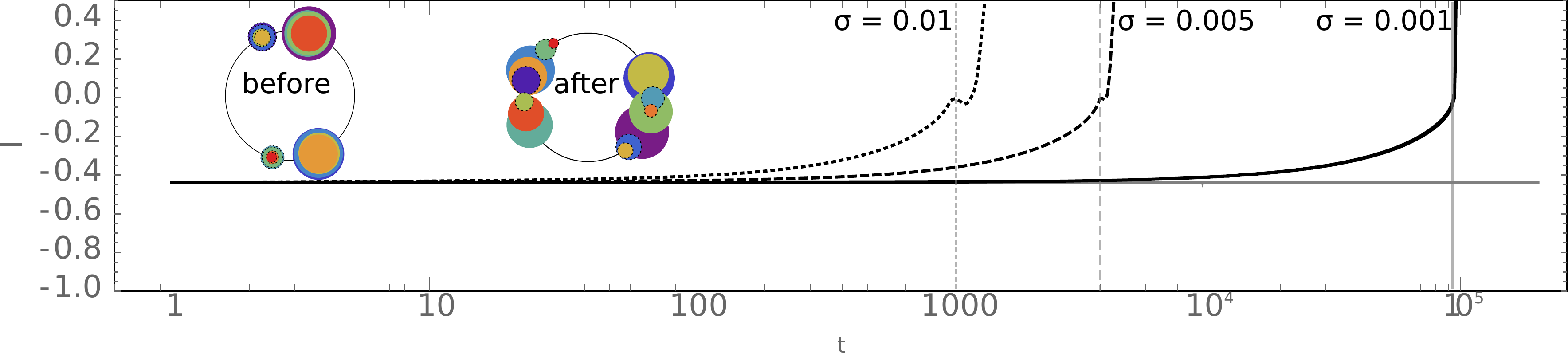}
		\end{center}
	\caption{Decay of the formerly ($\omega_i=\omega$) stable 4-cluster state, measured via the conserved quantity $I_{klmn}$ as defined in the text. The thick gray line shows the value of $I$ for ($\sigma$ = 0), black lines for $\sigma = 0.001$ (full), 0.005 (dashed), and 0.01 (dotted). The two insets show typical states of the system before and after the collision of the clusters.  The time axis is represented in a logarithmic scale to make the figure more readable.}
\label{fig11}
\end{figure}

A third type of long time scale is generated if we start with initial conditions of a four-cluster state, which satisfies the criteria for the Watanabe-Strogatz phenomenon and would be stable as long as $\sigma=0$.
The long decay time of the conserved quantity $I_{klmn}=\frac{\sin\frac{\varphi_k-\varphi_l}{2}\sin\frac{\varphi_m-\varphi_n}{2}}{
\sin\frac{\varphi_k-\varphi_m}{2}\sin\frac{\varphi_l-\varphi_n}{2}}$ as
described in \cite{watanabe} and predicted to be constant for $\omega_i=\omega$ in the four-cluster state, slowly decays.
In particular we show in Fig.~\ref{fig11} how long it takes the system to leave the 4-cluster state. The decay time depends on the width of the Gaussian distribution for regular $\omega$'s. When the value of $I(\sigma)$ starts to diverge from a constant $I(0)$, clusters move towards each other (while for $\sigma=0$ the cluster difference remains constant). Once the clusters collide, the system evolves to one of the available attractors. If the clusters containing oscillators $k$ and $l$ collide, or $m$ and $n$, $I$ becomes zero (right inset), defining the decay time. Alternatively, if the clusters containing $k$ and $m$ collide, or $l$ and $n$, $I$ becomes infinite. Again, in agreement with  Fig.~\ref{fig10}, the smaller the width of the distribution about the uniform case, the longer the transient of the system to reach the attractor.

As we shall show in forthcoming work, the variety of transient times, in particular long transients lead to physical aging of our deterministic oscillatory systems, as long as both time instants of the autocorrelation measurements fall into the transient period, so that stochastic fluctuations are not necessary for aging.

Basic ingredients for our observations were antagonistic couplings, leading to a  rich attractor space, together with a distribution of natural frequencies and an assignment to a regular grid.
Therefore we expect that our results are not restricted to our choice of model, but point to a mechanism, which may be  more generally at work in biological systems, where these basic ingredients are frequently found.
 \vskip1pt

Financial support from Deutsche Forschungsgemeinschaft (DFG, contract ME-1332/25-1) is gratefully acknowledged.

\end{document}